\documentclass[prodmode,acmtoms]{acmsmall}
\usepackage{graphicx}
\usepackage{natbib}
\usepackage[ruled]{algorithm2e}
\renewcommand{\algorithmcfname}{ALGORITHM}
\SetAlFnt{\small}
\SetAlCapFnt{\small}
\SetAlCapNameFnt{\small}
\SetAlCapHSkip{0pt}
\IncMargin{-\parindent}

\acmVolume{1}
\acmNumber{1}
\acmArticle{1}
\acmYear{2014}
\acmMonth{12}

\title{An implementation of a randomized algorithm\\
       for principal component analysis}
\author{
Arthur Szlam
\affil{Facebook Artificial Intelligence Research}
Yuval Kluger
\affil{Yale University Department of Pathology}
Mark Tygert
\affil{Facebook Artificial Intelligence Research and Yale University}
}

\begin{document}

\begin{abstract}
Recent years have witnessed intense development of randomized methods
for low-rank approximation. These methods target principal component analysis
(PCA) and the calculation of truncated singular value decompositions (SVD).
The present paper presents an essentially black-box, fool-proof implementation
for Mathworks' MATLAB, a popular software platform for numerical computation.
As illustrated via several tests, the randomized algorithms for low-rank
approximation outperform or at least match the classical techniques (such as
Lanczos iterations) in basically all respects: accuracy, computational
efficiency (both speed and memory usage), ease-of-use, parallelizability, and
reliability. However, the classical procedures remain the methods of choice for
estimating spectral norms, and are far superior for calculating the least
singular values and corresponding singular vectors (or singular subspaces).
\end{abstract}

\markboth{A.\ Szlam et al.}{An implementation of a randomized algorithm
                           for principal component analysis}
\category{Mathematics of Computing}{Mathematical Software}{Statistical Software}
\terms{Algorithms, Performance}
\keywords{principal component analysis, PCA, singular value decomposition, SVD}

\acmformat{Arthur Szlam, Yuval Kluger, and Mark Tygert, 2014. An implementation
of a randomized algorithm for principal component analysis.}

\begin{bottomstuff}
This work was supported in part by a US DoD DARPA Young Faculty Award
and US NIH Grant R0-1 CA158167.

Author's addresses:
A.\ Szlam, Facebook, 8th floor, 770 Broadway, New York, NY 10003;
Y.\ Kluger, Yale University, School of Medicine, Department of Pathology,
Suite 505F, 300 George St., New Haven, CT 06511;
M.\ Tygert, Facebook, 1 Hacker Way, Menlo Park, CA 94025
\end{bottomstuff}

\maketitle

\section{Introduction}
\label{intro}

Randomized algorithms for low-rank approximation in principal component
analysis and singular value decomposition have drawn a remarkable amount of
attention in recent years, as summarized in the review
of~\citet{halko-martinsson-tropp}. The present paper describes developments
that have led to an essentially black-box, fool-proof MATLAB implementation of
these methods, and benchmarks the implementation against the standards.
For applications to principal component analysis, the performance of the
randomized algorithms
run under their default parameter settings meets or exceeds (often exceeding
extravagantly) the standards'. In contrast to the existing standards, the
randomized methods are gratifyingly easy to use, rapidly and reliably producing
nearly optimal accuracy without extensive
tuning of parameters (in accordance with guarantees that rigorous proofs
provide). The present paper concerns implementations for MATLAB; a related
development is the C++ package ``libSkylark'' of~\citet{skylark}. Please beware
that the randomized methods on their own are ill-suited for calculating small
singular values and the corresponding singular vectors (or singular subspaces),
including ground states and corresponding energy levels of Hamiltonian
systems; the present article focuses on principal component analysis involving
low-rank approximations.

The present paper has the following structure:
Section~\ref{overview} outlines the randomized methods.
Section~\ref{nystrom} stabilizes an accelerated method for nonnegative-definite
self-adjoint matrices.
Section~\ref{accuracy} details subtleties involved in measuring the accuracy
of low-rank approximations.
Section~\ref{optimizations} tweaks the algorithms to improve the performance
of their implementations.
Section~\ref{proprob} discusses some issues with one of the most popular
existing software packages.
Section~\ref{perfdense} tests the different methods on dense matrices.
Section~\ref{perfsparse} tests the methods on sparse matrices.
Section~\ref{conclusion} draws several conclusions.

\section{Overview}
\label{overview}

The present section sketches randomized methods
for singular value decomposition (SVD) and
principal component analysis (PCA). The definitive treatment ---
that of~\citet{halko-martinsson-tropp} --- gives details,
complete with guarantees of superb accuracy;
see also the sharp analysis of~\citet{witten-candes} and the new work
of~\citet{woodruff} and others. PCA is the same as the SVD,
possibly after subtracting from each column its mean and otherwise
normalizing the columns (or rows) of the matrix being approximated.

PCA is most useful when the greatest singular values are reasonably greater
than those in the tail. Suppose that $k$ is a positive integer that is
substantially less than both dimensions of the matrix $A$ being analyzed,
such that the $k$ greatest singular values include the greatest singular values
of interest. The main output of PCA would then be a good rank-$k$ approximation
to $A$ (perhaps after centering or otherwise normalizing the columns or rows
of $A$). The linear span of the columns of this rank-$k$ approximation is an
approximation to the range of $A$. Given that $k$ is substantially less than
both dimensions of $A$, the approximate range is relatively low dimensional.

Shortly we will discuss an efficient construction of $k$ orthonormal vectors
that nearly span the range of $A$; such vectors enable the efficient
construction of an approximate SVD of $A$, as follows.
Denoting by $Q$ the matrix whose columns are these vectors,
the orthogonal projector on their span is $QQ^*$ (where $Q^*$ denotes
the adjoint --- the conjugate transpose --- of $Q$), and so
\begin{equation}
\label{i1}
A \approx QQ^* A,
\end{equation}
since this orthogonal projector nearly preserves the range of $A$. Because the
number $k$ of columns of $Q$ is substantially less than both dimensions of $A$,
we can efficiently compute
\begin{equation}
\label{i2}
B = Q^* A,
\end{equation}
which has only $k$ rows. We can then efficiently calculate an SVD
\begin{equation}
\label{i3}
W \Sigma V^* = B,
\end{equation}
where the columns of $W$ are orthonormal, as are the columns of $V$, and
$\Sigma$ is a diagonal $k \times k$ matrix whose entries are all nonnegative.
Constructing
\begin{equation}
\label{i4}
U = Q W
\end{equation}
and combining formulae~(\ref{i1})--(\ref{i4}) then yields the SVD
\begin{equation}
\label{approx}
U \Sigma V^* \approx A.
\end{equation}
If $k$ is substantially less than both dimensions of $A$,
then $A$ has far more entries than any other matrix in the above calculations.

Thus, provided that we can efficiently construct $k$ orthonormal vectors that
nearly span the range of $A$, we can efficiently construct an SVD that closely
approximates $A$ itself (say, in the sense that the spectral norm of the
difference between $A$ and the approximation to $A$ is small relative to the
spectral norm of $A$). In order to identify vectors in the range of $A$, we can
apply $A$ to random vectors --- after all, the result of applying $A$ to any
vector is a vector in the range of $A$. If we apply $A$ to $k$ random vectors,
then the results will nearly span the range of $A$, with extremely high
probability (and the probability of capturing most of the range is even higher
if we apply $A$ to a few extra random vectors).
Rigorous mathematical proofs (given by~\citet{halko-martinsson-tropp}
and by~\citet{witten-candes}, for example) show that the probability
of missing a substantial part of the range of $A$ is negligible, so long as
the vectors to which we apply $A$ are sufficiently random (which is so if,
for example, the entries of these vectors are independent and identically
distributed --- i.i.d.\ --- each drawn from a standard normal distribution).
Since the results of applying $A$ to these random vectors nearly span the range
of $A$, applying the Gram-Schmidt process (or other methods for constructing
QR decompositions) to these results yields an orthonormal basis
for the approximate range of $A$, as desired.

This construction is particularly efficient whenever $A$ can be applied
efficiently to arbitrary vectors, and is always easily parallelizable since the
required matrix-vector multiplications are independent of each other. The
construction of $B$ in formula~(\ref{i2}) requires further matrix-vector
multiplications, though there exist algorithms for the SVD of $A$ that avoid
explicitly constructing $B$ altogether, via the application of $A^*$ to random
vectors, identifying the range of $A^*$.
The full process is especially efficient when both $A$ and $A^*$ can be applied
efficiently to arbitrary vectors, and is always easily parallelizable.
Further accelerations are possible when $A$ is self-adjoint (and even more when
$A$ is nonnegative definite).

If the singular values of $A$ decay slowly, then the accuracy of the
approximation in formula~(\ref{approx}) may be lower than desired; the long
tail of singular values that we are trying to neglect may pollute the results
of applying $A$ to random vectors. To suppress this tail of singular values
relative to the singular values of interest (the leading $k$ are those of
interest), we may identify the range of $A$ by applying $AA^*A$ rather than $A$
itself --- the range of $AA^*A$ is the same as the range of $A$, yet the tail
of singular values of $AA^*A$ is lower (and decays faster) than the tail of
singular values of $A$, relative to the singular values of interest. Similarly,
we may attain even higher accuracy by applying $A$ and $A^*$ in succession to
each of the random vectors multiple times. The accuracy obtained thus
approaches the best possible exponentially fast, as proven
by~\citet{halko-martinsson-tropp}.

In practice, we renormalize after each application of $A$ or $A^*$, to avoid
problems due to floating-point issues such as roundoff or dynamic range
(overflow and underflow). The renormalized methods resemble the classical
subspace or power iterations (QR or LR iterations) widely used for spectral
computations, as reviewed by~\citet{golub-van_loan}. Our MATLAB codes ---
available at http://tygert.com/software.html ---  provide full details,
complementing the summary in Section~\ref{optimizations} below.

\section{Stabilizing the Nystr\"om method}
\label{nystrom}

Enhanced accuracy is available when the matrix $A$ being approximated
has special properties. For example, Algorithm~5.5 (the ``Nystr\"om method'')
of~\citet{halko-martinsson-tropp} proposes the following scheme
for processing a nonnegative-definite self-adjoint matrix $A$,
given a matrix $Q$ satisfying formula~(\ref{i1}), that is, $A \approx Q Q^* A$,
such that the columns of $Q$ are orthonormal:

Form
\begin{equation}
\label{start}
B_1 = A Q.
\end{equation}
Construct
\begin{equation}
\label{second}
B_2 = Q^* B_1.
\end{equation}
Compute a triangular matrix $C$ for the Cholesky decomposition
\begin{equation}
\label{Cholesky}
C^* C = B_2.
\end{equation}
By solving linear systems of equations, construct
\begin{equation}
\label{pseudo}
F = B_1 C^{-1} = ((C^*)^{-1} B_1^*)^*.
\end{equation}
Compute the SVD
\begin{equation}
U S V^* = F,
\end{equation}
where the columns of $U$ are orthonormal,
the columns of $V$ are also orthonormal,
$S$ is diagonal, and all entries of $S$ are nonnegative.
Set
\begin{equation}
\label{finish}
\Sigma = S^2.
\end{equation}
Then (as demonstrated by~\citet{halko-martinsson-tropp})
$A \approx U \Sigma U^*$, and the accuracy of this approximation should be
better than that obtained using formulae~(\ref{i2})--(\ref{i4}).

Unfortunately, this procedure can be numerically unstable. Even if $A$
is self-adjoint and nonnegative-definite, $B_2$ constructed in~(\ref{second})
may not be strictly positive definite (especially with roundoff errors),
as required for the Cholesky decomposition in~(\ref{Cholesky}).
To guarantee numerical stability, we need only replace the triangular matrix
for the Cholesky decomposition in~(\ref{Cholesky})
from formulae~(\ref{start})--(\ref{finish})
with the calculation of a self-adjoint square-root $C$ of $B_2$,
that is, with the calculation of a self-adjoint matrix $C$ such that
\begin{equation}
C^2 = B_2.
\end{equation}
The SVD of $B_2$ provides a convenient means
for computing the self-adjoint square-root $C$.
Technically, the inverse in formula~(\ref{pseudo}) should become
a (regularized) pseudoinverse, or, rather, should construct
backwardly stable solutions to the associated systems of linear equations.

Replacing the Cholesky factor with the self-adjoint square-root
is only one possibility for guaranteeing numerical stability.
Our MATLAB codes instead add (and subtract) an appropriate multiple
of the identity matrix to ensure strict positive definiteness
for the Cholesky decomposition. This alternative is somewhat more efficient,
but its implementation is more involved.
The simpler approach via the self-adjoint square-root should be sufficient
in most circumstances.

\section{Measuring accuracy}
\label{accuracy}

For most applications of principal component analysis, the spectral norm of the
discrepancy, $\| A - U \Sigma V^*\|$, where $U \Sigma V^*$ is the computed
approximation to $A$, is the most relevant measure of accuracy.
The spectral norm $\|H\|$ of a matrix $H$ is the maximum value of $|Hx|$,
maximized over every vector $x$ such that $|x| = 1$, where $|Hx|$ and $|x|$
denote the Euclidean norms of $Hx$ and $x$ (the spectral norm of $H$ is also
equal to the greatest singular value of $H$). The spectral norm is unitarily
invariant, meaning that its value is the same with respect to any unitary
transformation of the rows or any unitary transformation of the columns ---
that is, the value is the same with regard to any orthonormal basis of the
domain and to any orthonormal basis of the range or codomain
\citep{golub-van_loan}.

If the purpose of a principal component analysis is to facilitate dimension
reduction, denoising, or calibration, then the effectiveness of the
approximation at reconstructing the original matrix is the relevant metric for
measuring accuracy. This would favor measuring accuracy as the size of the
difference between the approximation and the matrix being approximated, as in
the spectral-norm discrepancy, rather than via direct assessment of the
accuracy of singular values and singular vectors. The spectral norm is a
uniformly continuous function of the matrix entries of the discrepancy,
unlike the relative accuracy of singular values (the relative accuracy of an
approximation $\tilde{\sigma}$ to a singular value $\sigma$ is
$|\tilde{\sigma}-\sigma|/\sigma$); continuity means that the spectral norm is
stable to small perturbations of the entries \citep{golub-van_loan}. 

The Frobenius norm of the difference between the approximation and the
matrix being approximated is unitarily invariant as is the spectral norm,
and measures the size of the discrepancy as does the spectral norm (the
Frobenius norm is the square root of the sum of the squares of the matrix
entries) \citep{golub-van_loan}. Even so, the spectral norm is generally
preferable for big data subject to noise. Noise often manifests as a long tail
of singular values which individually are much smaller than the leading
singular values but whose total energy may approach or even exceed the leading
singular values'. For example, the singular values for a signal corrupted by
white noise flatten out sufficiently far out in the tail \citep{glossary}.
The sum of the squares of the singular values corresponding to white noise
or to pink noise diverges when adding further singular values as the dimensions
of the matrix increase \citep{glossary}. The square root of the sum of
the squares of the singular values in the tail thus overwhelms
the leading singular values for big matrices subject to white or pink noise
(as well as for other types of noise). Such noise can mask the contribution of
the leading singular values to the Frobenius norm (that is, to the square root
of the sum of squares); the ``signal'' has little effect on the Frobenius norm,
as this norm depends almost entirely on the singular values corresponding
to ``noise.''

Since the Frobenius norm is the square root of the sum of the squares
of {\it all} entries, the Frobenius norm throws together all the noise
from all directions. Of course, noise afflicts the spectral norm, too,
but only the noise in one direction at a time --- noise from noisy directions
does not corrupt a direction that has a high signal-to-noise ratio.
The spectral norm can detect and extract a signal so long as the singular
values corresponding to signal are greater than each of the singular values
corresponding to noise; in contrast, the Frobenius norm can detect the signal
only when the singular values corresponding to signal are greater than the
square root of the {\it sum} of the squares of {\it all} singular values
corresponding to noise. Whereas the individual singular values may not get
substantially larger as the dimensions of the matrix increase, the sum
of the squares may become troublingly large in the presence of noise.
With big data, noise may overwhelm the Frobenius norm. In the words
of Joel A. Tropp, Frobenius-norm accuracy may be ``vacuous'' in a noisy
environment. The spectral norm is comparatively robust to noise.

To summarize, a long tail of singular values that correspond to noise or are
otherwise unsuitable for designation as constituents of the ``signal'' can
obscure the signal of interest in the leading singular values and singular
vectors, when measuring accuracy via the Frobenius norm. Principal component
analysis is most useful when retaining only the leading singular values and
singular vectors, and the spectral norm is then more informative than the
Frobenius norm.

Fortunately, estimating the spectral norm is straightforward and reliable
using the power method with a random starting vector. Theorem 4.1(a)
of~\citet{kuczynski-wozniakowski} proves that the computed estimate lies
within a factor of two of the exact norm with overwhelmingly high probability,
and the probability approaches 1 exponentially fast as the number
of iterations increases. The guaranteed lower bound on the probability of
success is independent of the structure of the spectrum; the bound is highly
favorable even if there are no gaps between the singular values. Estimating the
spectral-norm discrepancy via the randomized power method is simple, reliable,
and highly informative.

\section{Algorithmic optimizations}
\label{optimizations}

The present section describes several improvements effected in our MATLAB codes
beyond the recommendations of~\citet{halko-martinsson-tropp}.

As~\citet{shabat-shmueli-averbuch} observed, computing the LU decomposition is
typically more efficient than computing the QR decomposition, and both are
sufficient for most stages of the randomized algorithms for low-rank
approximation. Our MATLAB codes use LU decompositions whenever possible. For
example, given some number of iterations, say its $= 4$, and given an
$n \times k$ random matrix $Q$, the core iterations in the case of a
self-adjoint $n \times n$ matrix $A$ are

{\tt

\begin{tabbing}

\quad \= \quad \= \quad \= \quad \\

for it = 1:its \\\\

\> Q = A*Q; \\\\

\> if(it < its) \\

\> \> [Q,R] = lu(Q); \\

\> end \\\\

\> if(it == its) \\

\> \> [Q,R,E] = qr(Q,0); \\
 
\> end \\\\

end \\

\end{tabbing}

}

\noindent In all but the last of these iterations, an LU decomposition
renormalizes $Q$ after the multiplication with $A$. In the last iteration, a
pivoted
QR decomposition renormalizes $Q$, ensuring that the columns of the resulting
$Q$ are orthonormal. Incidentally, since the initial matrix $Q$ was random,
pivoting in the QR decomposition is not necessary;
replacing the line ``{\tt [Q,R,E] = qr(Q,0)}'' with ``{\tt [Q,R] = qr(Q,0)}''
sacrifices little in the way of numerical stability.

A little care in the implementation ensures that the same code can efficiently
handle both dense and sparse matrices. For example, if $c$ is the $1 \times n$
vector whose entries are the means of the entries in the columns of an
$m \times n$ matrix $A$, then the MATLAB code {\tt Q = A*Q - ones(m,1)*(c*Q)}
applies the mean-centered $A$ to $Q$, without ever forming all entries of the
mean-centered $A$. Similarly, the MATLAB code {\tt Q = (Q\char'15{}*A)\char'15}
applies the adjoint of $A$ to $Q$, without ever forming the adjoint of $A$
explicitly, while taking full advantage of the storage scheme for $A$
(column-major ordering, for example).

Since the algorithms are robust to the quality of the random numbers used,
we can use the fastest available pseudorandom generators, for instance,
drawing from the uniform distribution over the interval $[-1,1]$
rather than from the normal distribution used in many theoretical analyses.

Another possible optimization is to renormalize only in the odd-numbered
iterations (that is, when the variable ``{\tt it}'' is odd in the above MATLAB
code). This particular acceleration would sacrifice accuracy. However,
as~\citet{rachakonda-silva-liu-calhoun} observed, this strategy can halve the
number of disk accesses/seeks required to process a matrix $A$ stored on
disk when $A$ is not self-adjoint. As our MATLAB codes do not directly support
out-of-core calculations, we did not incorporate this additional acceleration,
preferring the slightly enhanced accuracy of our codes.

\section{Hard problems for PROPACK}
\label{proprob}

PROPACK is a suite of software that can calculate low-rank approximations via
remarkable, intricate Lanczos methods, developed by~\citet{larsen}.
Unfortunately, PROPACK can be unreliable for computing low-rank approximations.
For example, using PROPACK's principal routine, ``lansvd,'' under its default
settings to process the diagonal matrix whose first three diagonal entries are
all 1, whose fourth through twentieth diagonal entries are all .999, and whose
other entries are all 0 yields the following wildly incorrect estimates for
the singular values 1, .999, and 0:

\smallskip

\noindent rank-20 approximation to a 30 $\times$ 30 matrix:
    1.3718, 1.3593, 1.3386, 1.3099, 1.2733, 1.2293, 1.1780, 1.1201, 1.0560,
    1.0000, 1.0000, 1.0000, 0.9990, 0.9990, 0.9990, 0.9990, 0.9990, 0.9990,
    0.9990, 0.9990 (all values should be 1 or .999)

\smallskip

\noindent rank-21 (with similar results for higher rank) approximation
to a 30 $\times$ 30 matrix:
    1.7884, 1.7672, 1.7321, 1.6833, 1.6213, 1.5466, 1.4599, 1.3619, 1.2537,
    1.1361, 1.0104, 1.0000, 1.0000, 1.0000, 0.9990, 0.9990, 0.9990, 0.9990,
    0.9990, 0.9990, 0.9990 (this last value should be 0; all others should be
    1 or .999)

\smallskip

\noindent rank-50 approximation to a 100 $\times$ 100 matrix:
    1.3437, 1.3431, 1.3422, 1.3409, 1.3392, 1.3372, 1.3348, 1.3321, 1.3289,
    1.3255, 1.3216, 1.3174, 1.3128, 1.3079, 1.3027, 1.2970, 1.2910, 1.2847,
    1.2781, 1.2710, 1.2637, 1.2560, 1.2480, 1.2397, 1.2310, 1.2220, 1.2127,
    1.2031, 1.1932, 1.1829, 1.1724, 1.1616, 1.1505, 1.1391, 1.1274, 1.1155,
    1.1033, 1.0908, 1.0781, 1.0651, 1.0519, 1.0385, 1.0248, 1.0110, 1.0000,
    1.0000, 1.0000, 0.9990, 0.9990, 0.9990 (the last 30 values should be 0;
    all others should be 1 or .999)

\smallskip

Full reorthogonalization fixes this --- as Rasmus Larsen,
the author of PROPACK, communicated to us ---
but at potentially great cost in computational efficiency.

\section{Performance for dense matrices}
\label{perfdense}

We calculate rank-$k$ approximations to an $m \times n$ matrix $A$ constructed
with specified singular values and singular vectors; $A = U \Sigma V^*$, with
$U$, $\Sigma$, and $V$ constructed thus: We specify the matrix $U$ of
left singular vectors to be the result of orthonormalizing via the Gram-Schmidt
process (or via an equivalent using QR-decompositions) $m$ vectors, each of
length $m$, whose entries are i.i.d.\ Gaussian random variables of zero mean
and unit variance. Similarly, we specify the matrix $V$ of right singular
vectors to be the result of orthonormalizing $n$ vectors, each of length $n$,
whose entries are i.i.d.\ Gaussian random variables of zero mean and unit
variance. We specify $\Sigma$ to be the $m \times n$ matrix whose entries off
the main diagonal are all zeros and whose diagonal entries are the singular
values $\sigma_1$,~$\sigma_2$, \dots, $\sigma_{\min(m,n)}$. We consider six
different distributions of singular values, testing each for two settings of
$m$ and $n$ (namely $m=n=1000$ and $m=100$, $n=200$). The first five
distributions of singular values are
\begin{equation}
\sigma_j = 1/j, \quad j = 1, 2, \dots, \min(m,n)
\end{equation}
\begin{equation}
\sigma_j = \left\{ \begin{array}{ll}
           1, & j = 1 \\
           2 \cdot 10^{-5}, & j = 2, 3, \dots, k \\
           10^{-5} \cdot (k+1)/j, & j = k+1, k+2, \dots, \min(m,n)
           \end{array} \right.
\end{equation}
\begin{equation}
\sigma_j = \left\{ \begin{array}{ll} 
           10^{-5(j-1)/(k-1)}, & j = 1, 2, \dots, k \\
           10^{-5} \cdot (k+1)/j, & j = k+1, k+2, \dots, \min(m,n)
           \end{array} \right.
\end{equation}
\begin{equation}
\sigma_j = \left\{ \begin{array}{ll}
           10^{-5(j-1)/(k-1)}, & j = 1, 2, \dots, k \\
           10^{-5}, & j = k+1 \\
           0, & j = k+2, k+3, \dots, \min(m,n)
           \end{array} \right.
\end{equation}
\begin{equation}
\sigma_j = \left\{ \begin{array}{ll}
           10^{-5} + (1-10^{-5}) \cdot (k-j)/(k-1), & j = 1, 2, \dots, k \\
           10^{-5} \cdot \sqrt{(k+1)/j}, & j = k+1, k+2, \dots, \min(m,n)
           \end{array} \right.
\end{equation}
The spectral norm of $A$ is 1 and the spectral norm of the difference between
$A$ and its best rank-$k$ approximation is $10^{-5}$ for each of the four
preceding examples. For the sixth example, we use for $\sigma_1$,~$\sigma_2$,
\dots, $\sigma_{\min(m,n)}$ the absolute values of $\min(m,n)$ i.i.d.\ Gaussian
random variables of zero mean and unit variance.

\begin{figure}
\centering
\parbox{.49\textwidth}{\includegraphics[width=.48\textwidth]{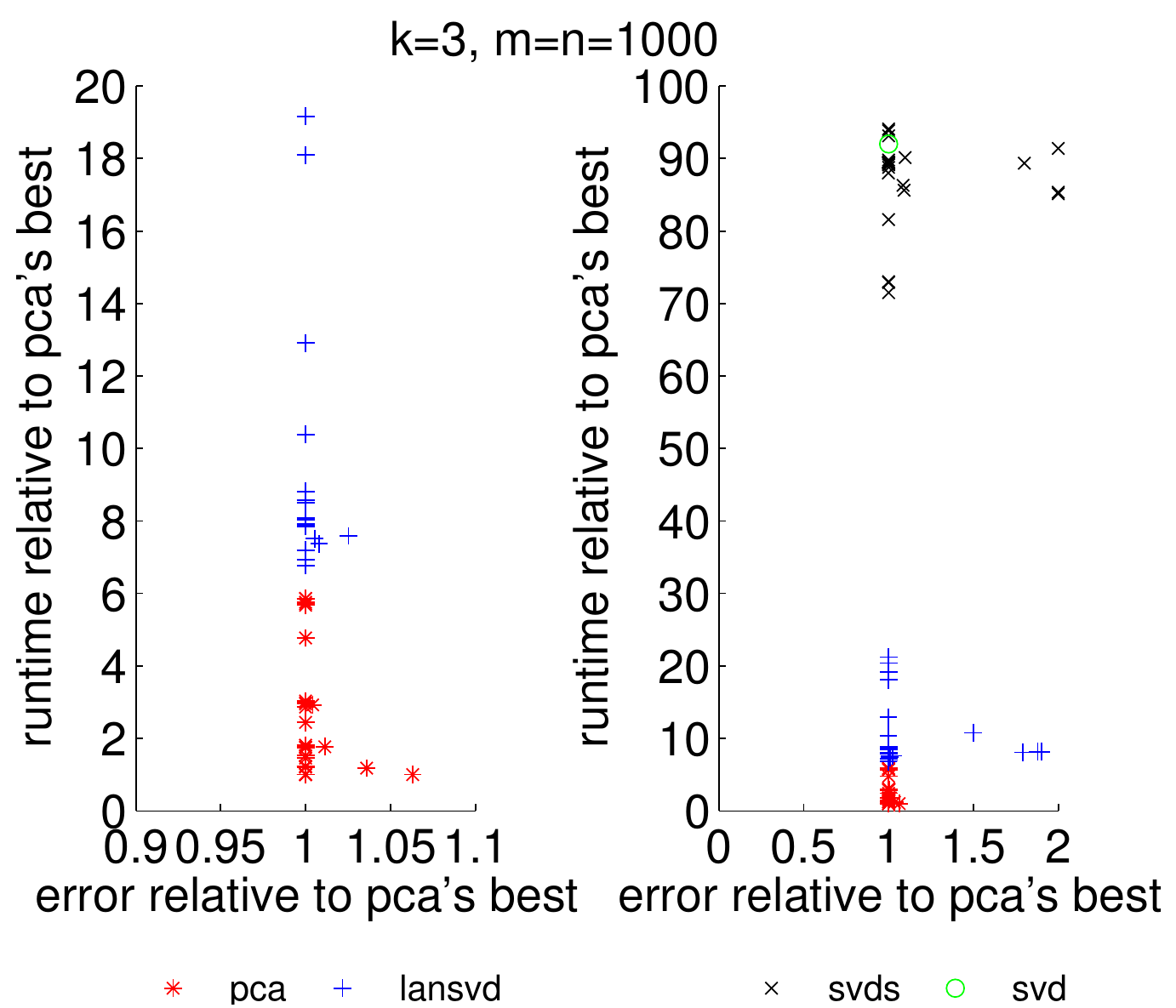}}
\parbox{.49\textwidth}{\includegraphics[width=.48\textwidth]{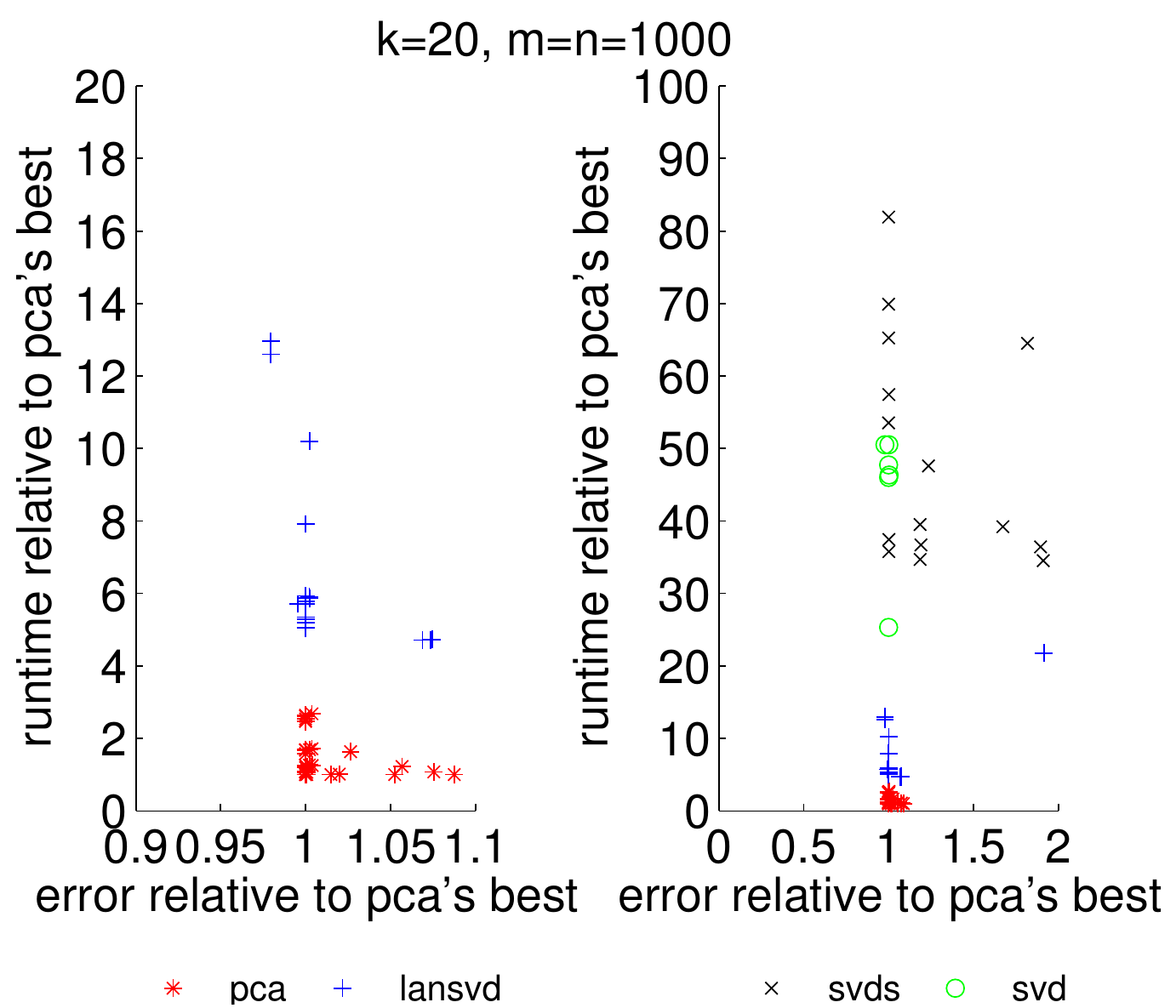}}
\\\vspace{.02\textwidth}
\parbox{.49\textwidth}{\includegraphics[width=.48\textwidth]{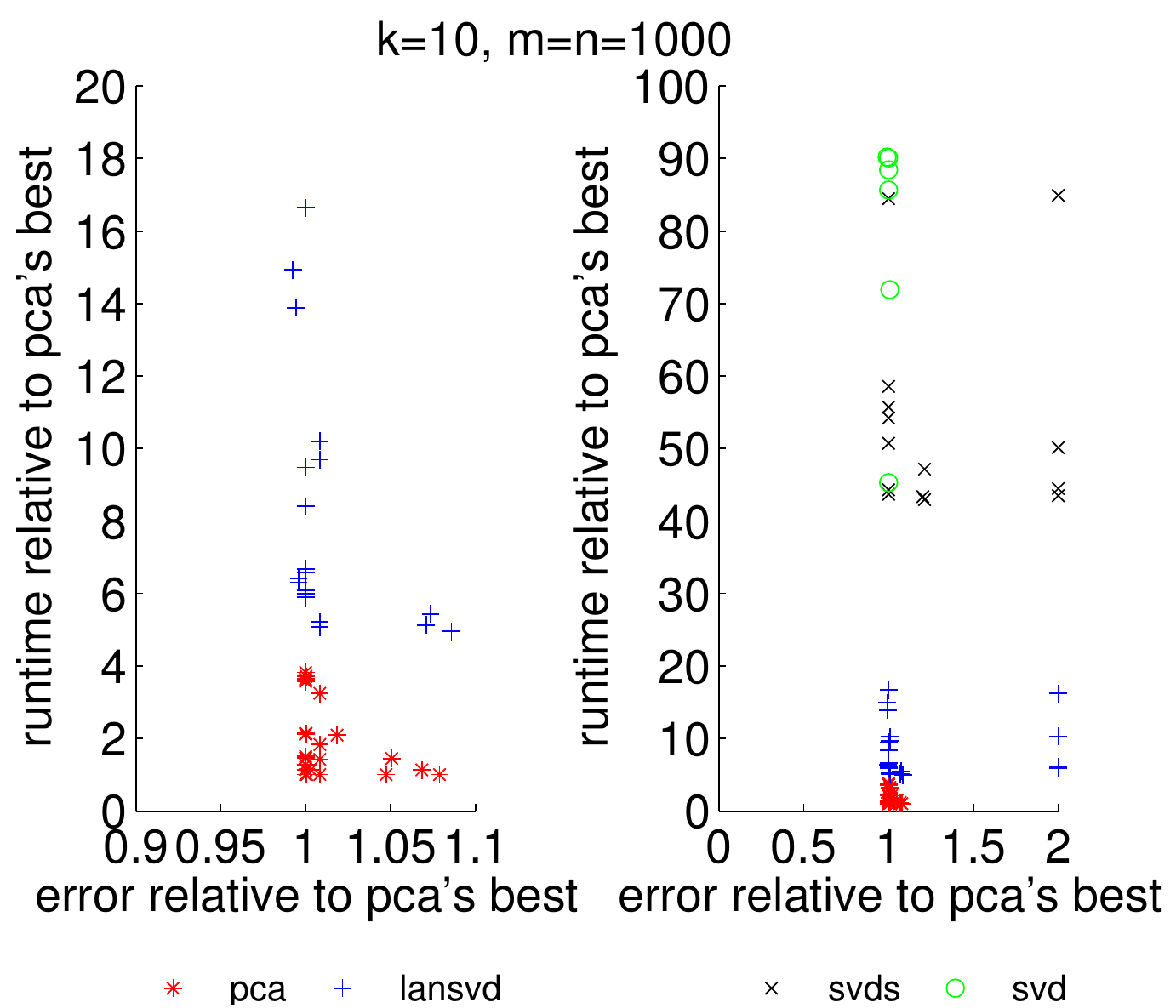}}
\parbox{.49\textwidth}{\includegraphics[width=.48\textwidth]{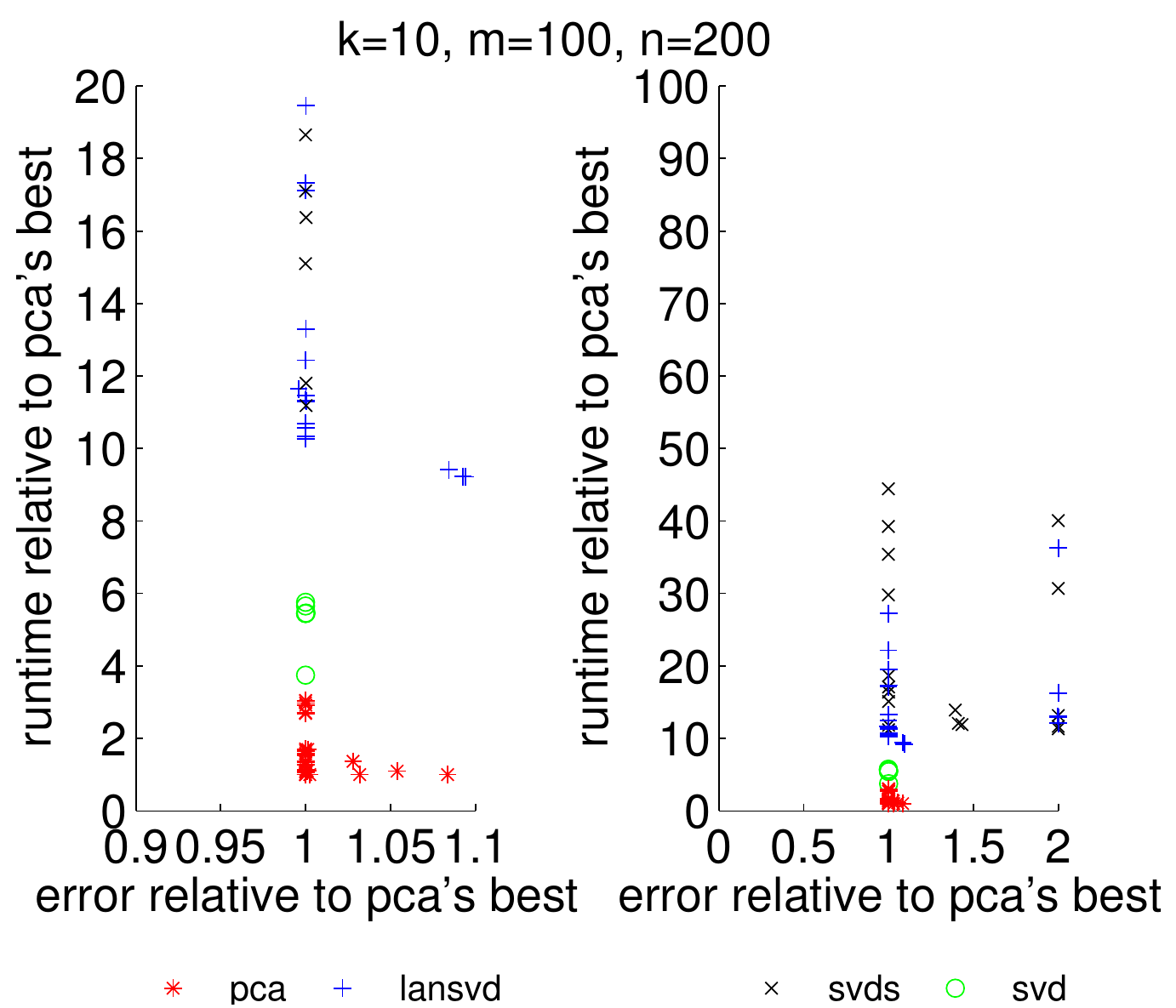}}
\\\ \\
\caption{Results for dense matrices}
\label{densefig}
\end{figure}

For each of the four parameter settings displayed in Figure~\ref{densefig}
(namely, $k=3$, $m=n=1000$; $k=10$, $m=n=1000$; $k=20$, $m=n=1000$; and $k=10$,
$m=100$, $n=200$), we plot the spectral-norm errors and runtimes for pca (our
code), lansvd (PROPACK of~\citet{larsen}), MATLAB's built-in svds (ARPACK
of~\citet{arpack}), and MATLAB's built-in svd (LAPACK of~\citet{lapack}). For
pca, we vary the oversampling parameter $l$ that specifies the number of random
vectors whose entries are i.i.d.\ as $l = k+2$,~$k+4$, $k+8$, $k+16$,~$k+32$;
we leave the parameter specifying the number of iterations at the default,
its $= 2$. For lansvd and svds, we vary the tolerance for convergence as tol
$= 10^{-8}$,~$10^{-4}$, 1, $10^4$,~$10^8$, capping the maximum number of
iterations to be $k$ --- the minimum possible --- when tol $= 10^8$. Each
plotted point represents the averages over ten randomized trials (the plots
look similar, but slightly busier, without any averaging). The red asterisks
correspond to pca, the blue ``plus signs'' correspond to lansvd, the black
``times signs'' correspond to svds, and the green circles correspond to svd.
Clearly pca reliably yields substantially higher performance. Please note that
lansvd is the closest competitor to pca, yet may return entirely erroneous
results without warning, as indicated in Section~\ref{proprob}.

For reference, the rank of the approximation being constructed is $k$,
and the matrix $A$ being approximated is $m \times n$. The plotted accuracy is
the spectral norm of the difference between $A$ and the computed rank-$k$
approximation. Each plot in Figure~\ref{densefig} appears twice,
with different ranges for the axes.

\begin{figure}
\centering
\includegraphics[width=.48\textwidth]{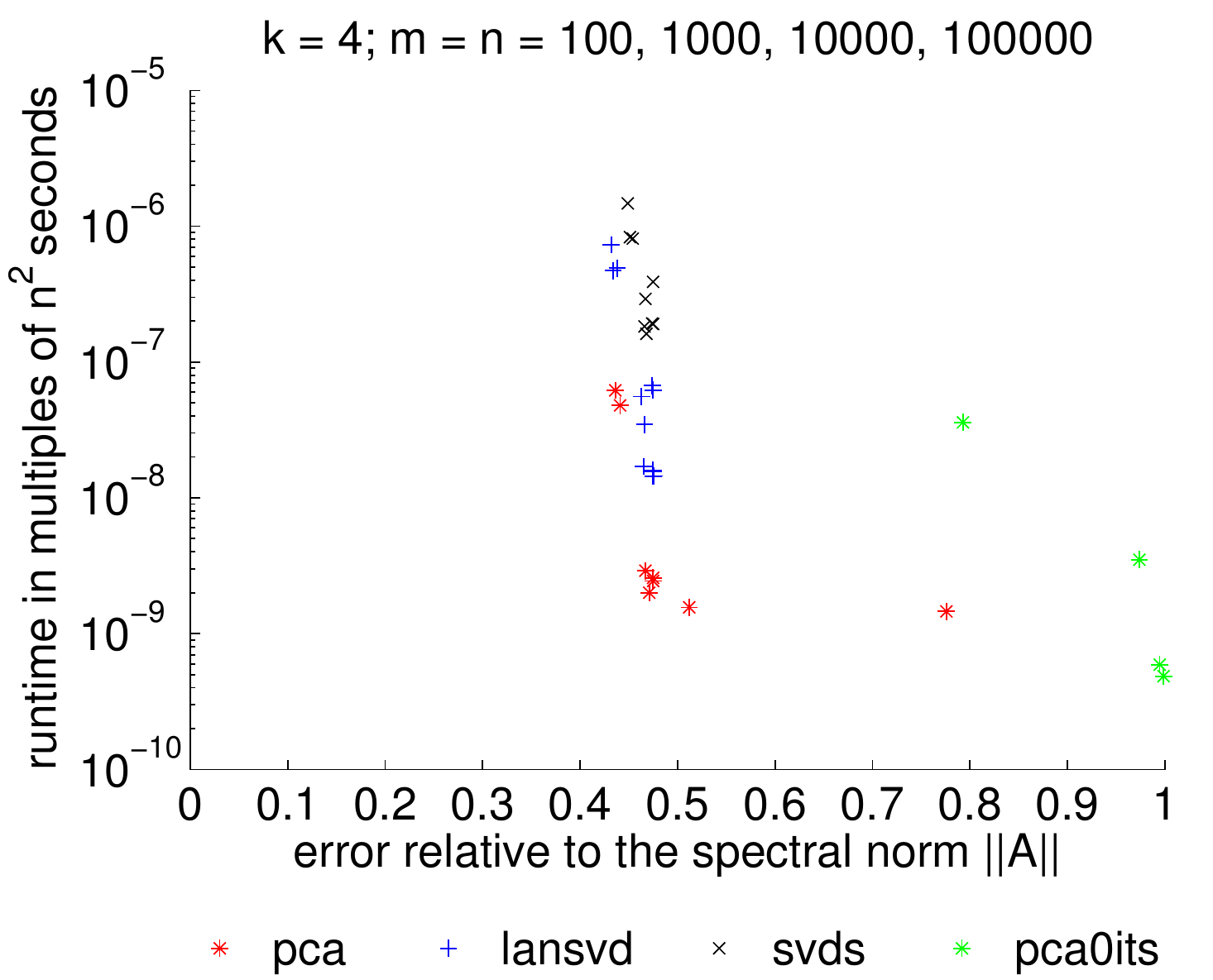}
\caption{Importance of power/subspace iterations}
\label{0its}
\end{figure}

We also construct rank-4 approximations to an $n \times n$ matrix $A$ whose
entries are i.i.d.\ Gaussian random variables of mean $\sqrt{30/n}$ and
variance 1, flipping the sign of the entry in row $i$ and column $j$
if $i \cdot j$ is odd. Such a matrix has two singular values that are roughly
twice as large as the largest of the others; without flipping the signs of the
entries, there would be only one singular value substantially larger than the
others, still producing results analogous to those reported below. For the four
settings $m = n = 100$,~$1000$, $10000$,~$100000$, Figure~\ref{0its} plots the
spectral-norm errors and runtimes for pca (our code), lansvd (PROPACK
of~\citet{larsen}), and MATLAB's built-in svds (ARPACK of~\citet{arpack}). For
pca, we use 0, 2, and 4 extra power/subspace iterations (setting its $= 0$, 2,
4 in our MATLAB codes), and plot the case of 0 extra iterations separately, as
pca0its; we leave the oversampling parameter $l$ specifying the number of
random vectors whose entries are i.i.d.\ at the default, $l = k+2$. For lansvd
and svds, we vary the tolerance for convergence as tol $= 10^{-2}$, 1, $10^8$,
capping the maximum number of iterations to be 4 --- the minimum possible ---
when tol $= 10^8$. The best possible spectral-norm accuracy of a rank-4
approximation is about half the spectral norm $\|A\|$ (not terribly accurate,
yet unmistakably more accurate than approximating $A$ by, say, a matrix whose
entries are all 0). Each plotted point represents the averages over ten
randomized trials (the plots look similar, but somewhat busier, without this
averaging). The red asterisks correspond to pca with its $= 2$ or its $= 4$,
the blue ``plus signs'' correspond to lansvd, the black ``times signs''
correspond to svds, and the green asterisks correspond to pca with its $= 0$,
that is, to pca0its. The plots omit svds for $m = n = 100000$, since the
available memory was insufficient for running MATLAB's built-in implementation.
Clearly pca is far more efficient. Some extra power/subspace iterations are
necessary to yield good accuracy; except for $m = n = 100000$, using only
its $= 2$ extra power/subspace iterations yields very nearly optimal accuracy,
whereas pca0its (pca with its $= 0$) produces very inaccurate approximations.

The computations used MATLAB 8.3.0.532 (R2014a) on a four-processor machine,
with each processor being an Intel Xeon E5-2680 v2 containing 10 cores,
where each core operated at 2.8\,GHz, with 25.6\,MB of L2 cache.
We did not consider the MATLAB Statistics Toolbox's own ``pca,'' ``princomp,''
and ``pcacov,'' as these compute all singular values and singular vectors
(not only those relevant for low-rank approximation), just like MATLAB's
built-in svd (in fact, these other functions call svd).

\section{Performance for sparse matrices}
\label{perfsparse}

We compute rank-$k$ approximations to each real $m \times n$ matrix $A$ from
the University of Florida sparse matrix collection of~\citet{davis-hu} with
$200 < m < 2000000$ and $200 < n < 2000000$, such that the original collection
provides no right-hand side for use in solving a linear system with $A$
(matrices for use in solving a linear system tend to be so well-conditioned
that forming low-rank approximations makes no sense).

\begin{figure}
\centering
\parbox{.49\textwidth}{\includegraphics[width=.48\textwidth]{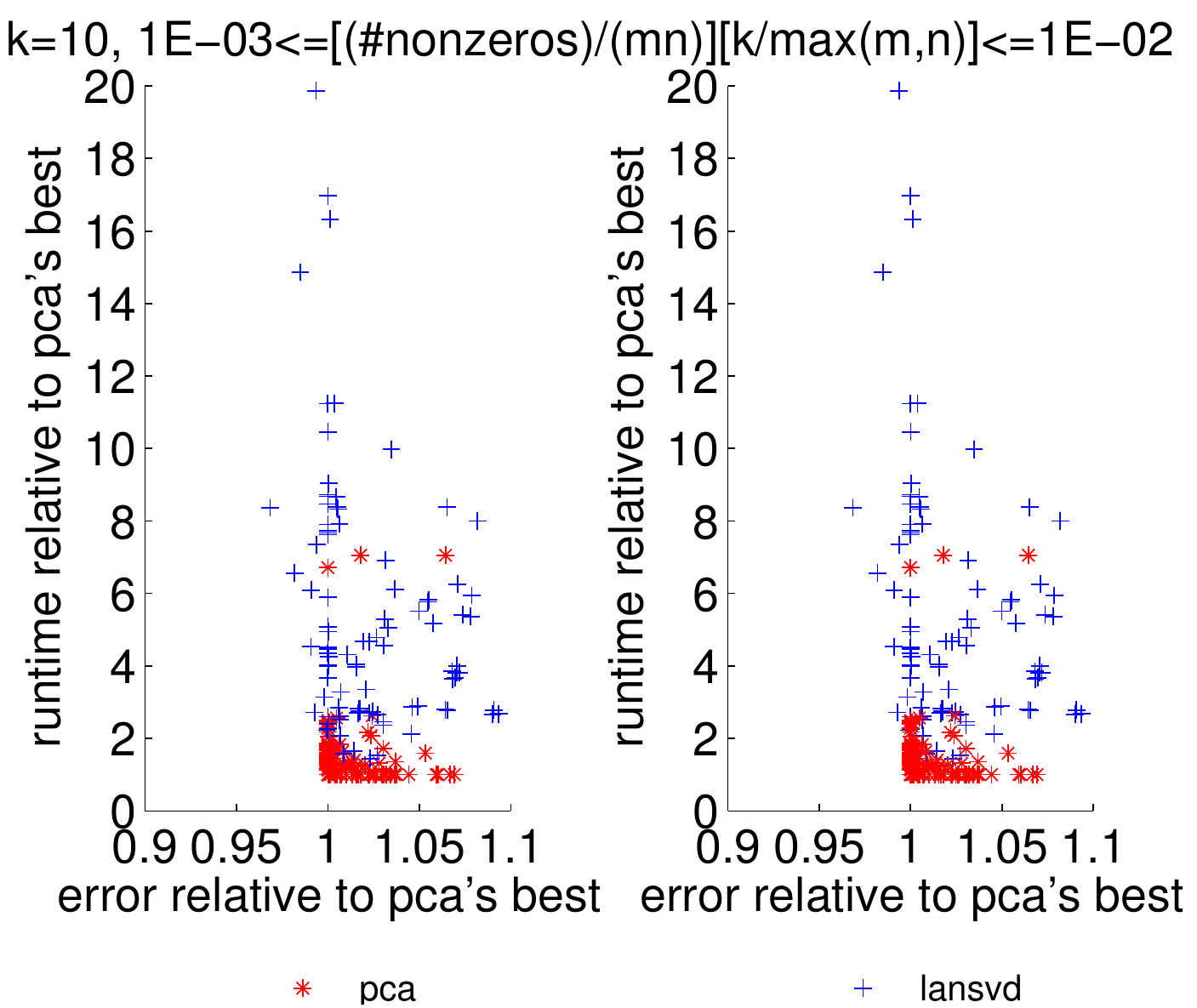}}
\parbox{.49\textwidth}{\includegraphics[width=.48\textwidth]{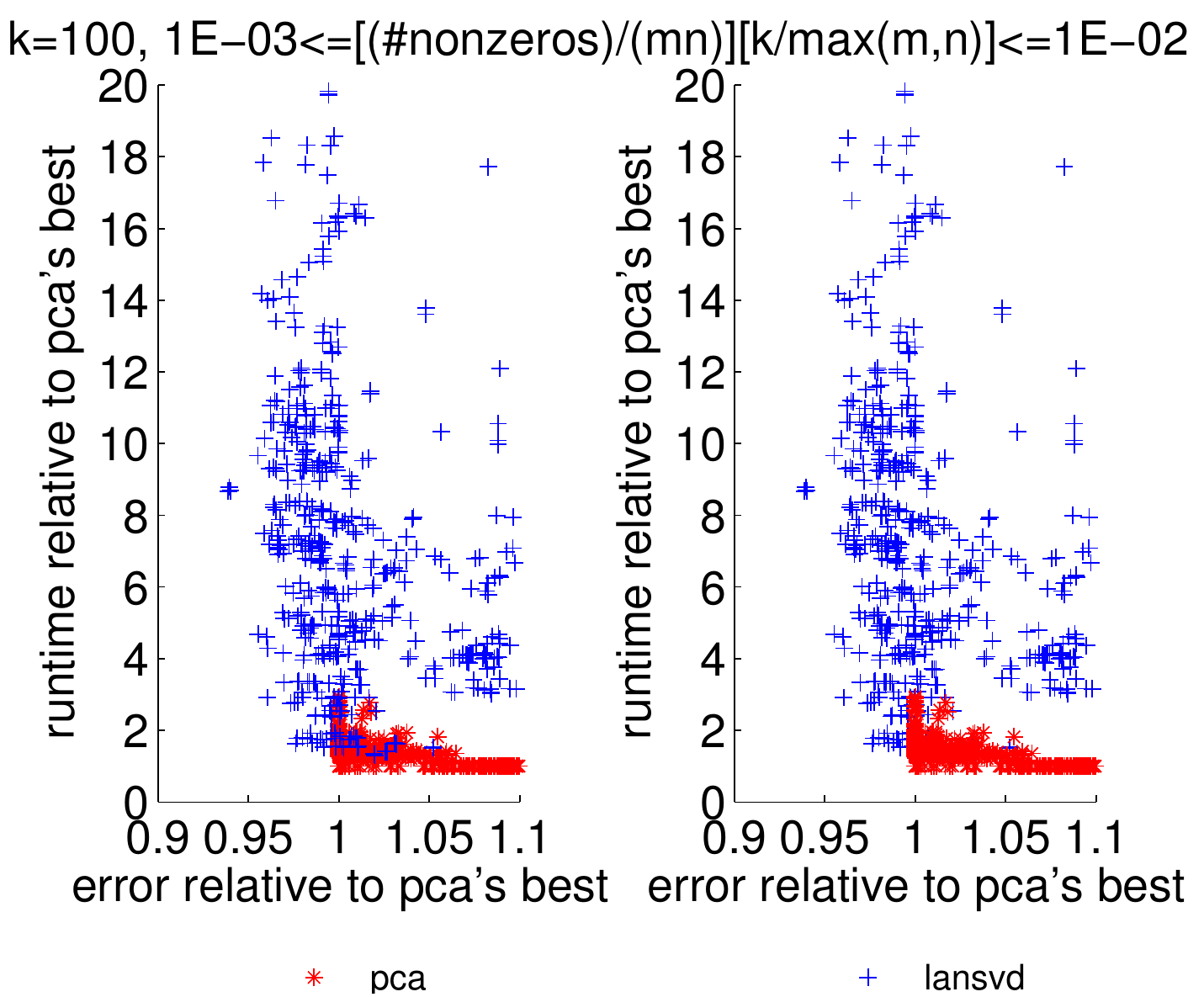}}
\\\vspace{.02\textwidth}
\parbox{.49\textwidth}{\includegraphics[width=.48\textwidth]{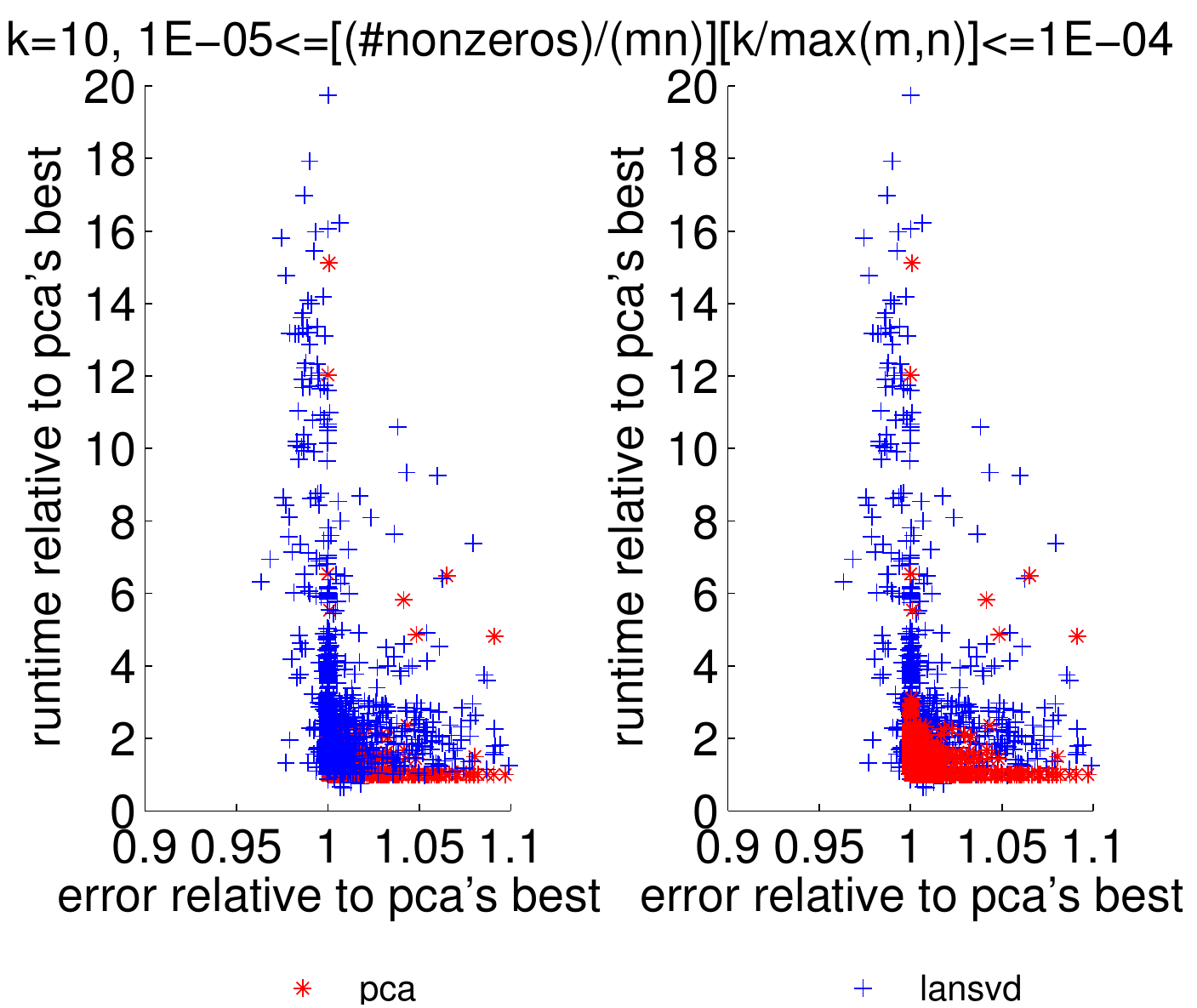}}
\parbox{.49\textwidth}{\includegraphics[width=.48\textwidth]{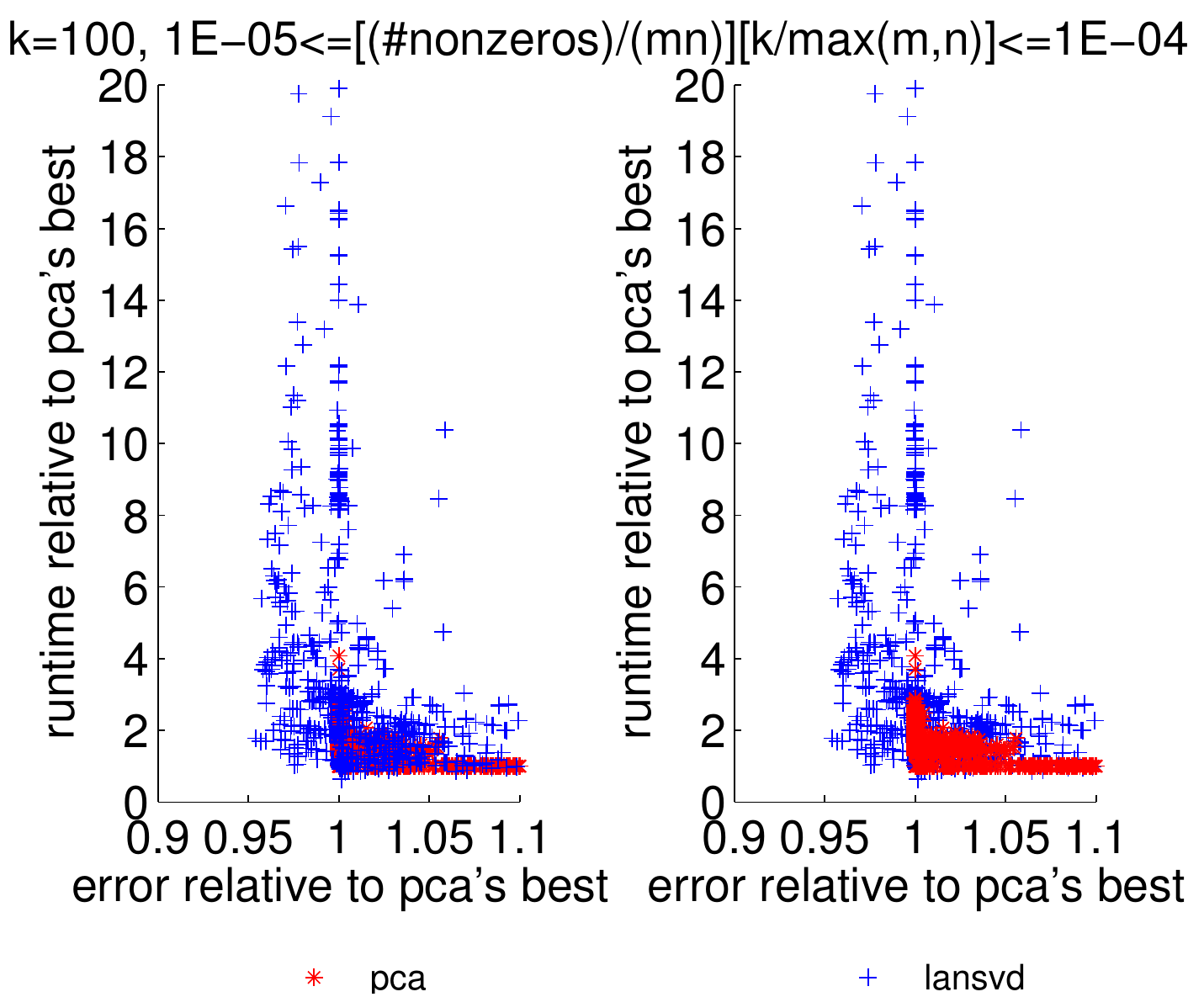}}
\\\vspace{.02\textwidth}
\parbox{.49\textwidth}{\includegraphics[width=.48\textwidth]{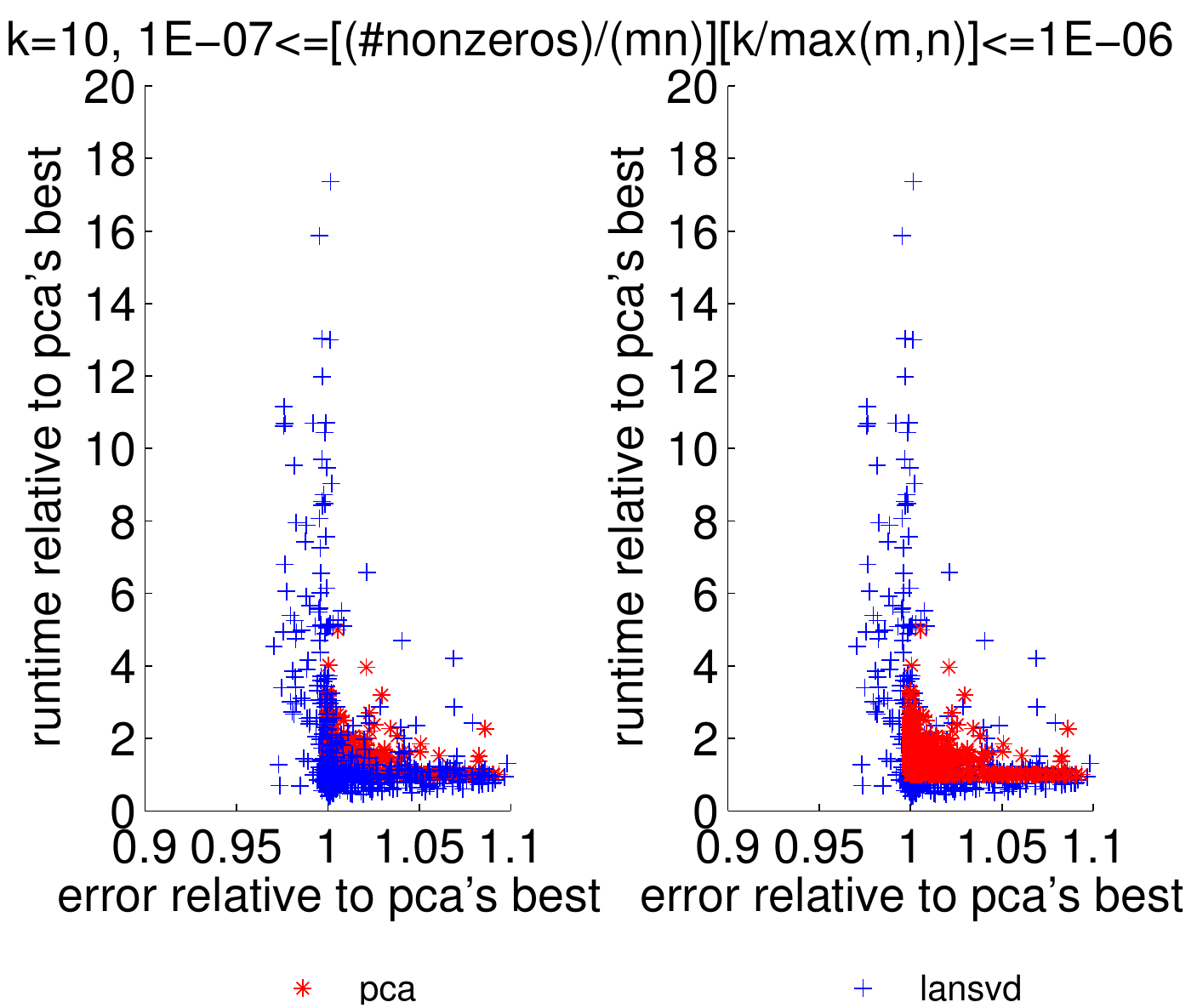}}
\parbox{.49\textwidth}{\includegraphics[width=.48\textwidth]{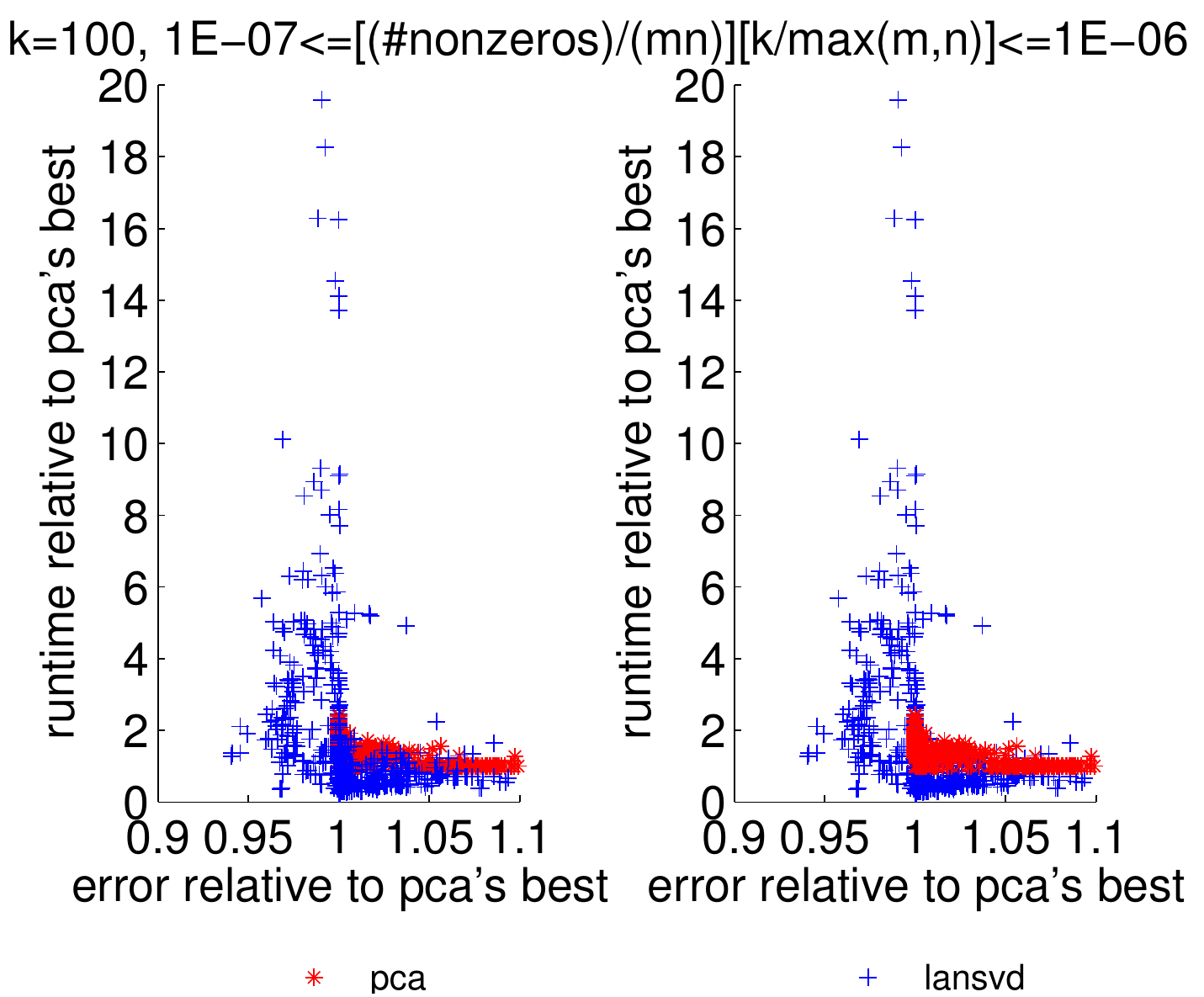}}
\\\ \\
\caption{pca vs.\ lansvd for sparse matrices}
\label{sparsefig}
\end{figure}

For each of the six parameter settings displayed in Figure~\ref{sparsefig}
(these settings are $10^{-3} \le \alpha \le 10^{-2}$,
as well as $10^{-5} \le \alpha \le 10^{-4}$
and $10^{-7} \le \alpha \le 10^{-6}$, for both $k=10$ and $k=100$, where
$\alpha = [\hbox{(number of nonzeros)}/(mn)]\cdot[k/\hbox{max}(m,n)]$),
we plot the spectral-norm errors and runtimes
for pca (our code) and lansvd (PROPACK of~\citet{larsen}). For pca, we vary the
parameter specifying the number of iterations as its $= 2$, 5, 8;
we leave the oversampling parameter $l$ that specifies the number of random
vectors whose entries are i.i.d.\ at the default, $l = k+2$. For lansvd, we
vary the tolerance for convergence as tol $= 10^{-5}$, 1, $10^3$.
The red asterisks correspond to pca and the blue ``plus signs'' correspond
to lansvd. Please note that lansvd may return entirely erroneous results
without warning, as indicated in Section~\ref{proprob}.

For reference, the rank of the approximation being constructed is $k$,
and the matrix $A$ being approximated is $m \times n$. The plotted accuracy is
the spectral norm of the difference between $A$ and the computed rank-$k$
approximation; pca's error was never greater than twice the best
for either algorithm for any setting of parameters.
Each plot in Figure~\ref{sparsefig} appears twice,
once with lansvd on top of pca, and once with pca on top of lansvd.
Figure~\ref{sparsefig} indicates that neither pca nor lansvd
is uniformly superior for sparse matrices.

We also use MATLAB's built-in svds to compute rank-$k$ approximations
to each real $m \times n$ matrix $A$ from the University of Florida collection
of~\citet{davis-hu} with $200 < m < 20000$ and $200 < n < 20000$,
such that the original collection provides no right-hand side for use
in solving a linear system with $A$ (matrices for use in solving
a linear system tend to be so well-conditioned that forming
low-rank approximations makes no sense). We report this additional test since
there was insufficient memory for running svds on the larger sparse matrices,
so we could not include results for svds in Figure~\ref{sparsefig}.

\begin{figure}
\centering
\parbox{.49\textwidth}{\includegraphics[width=.48\textwidth]{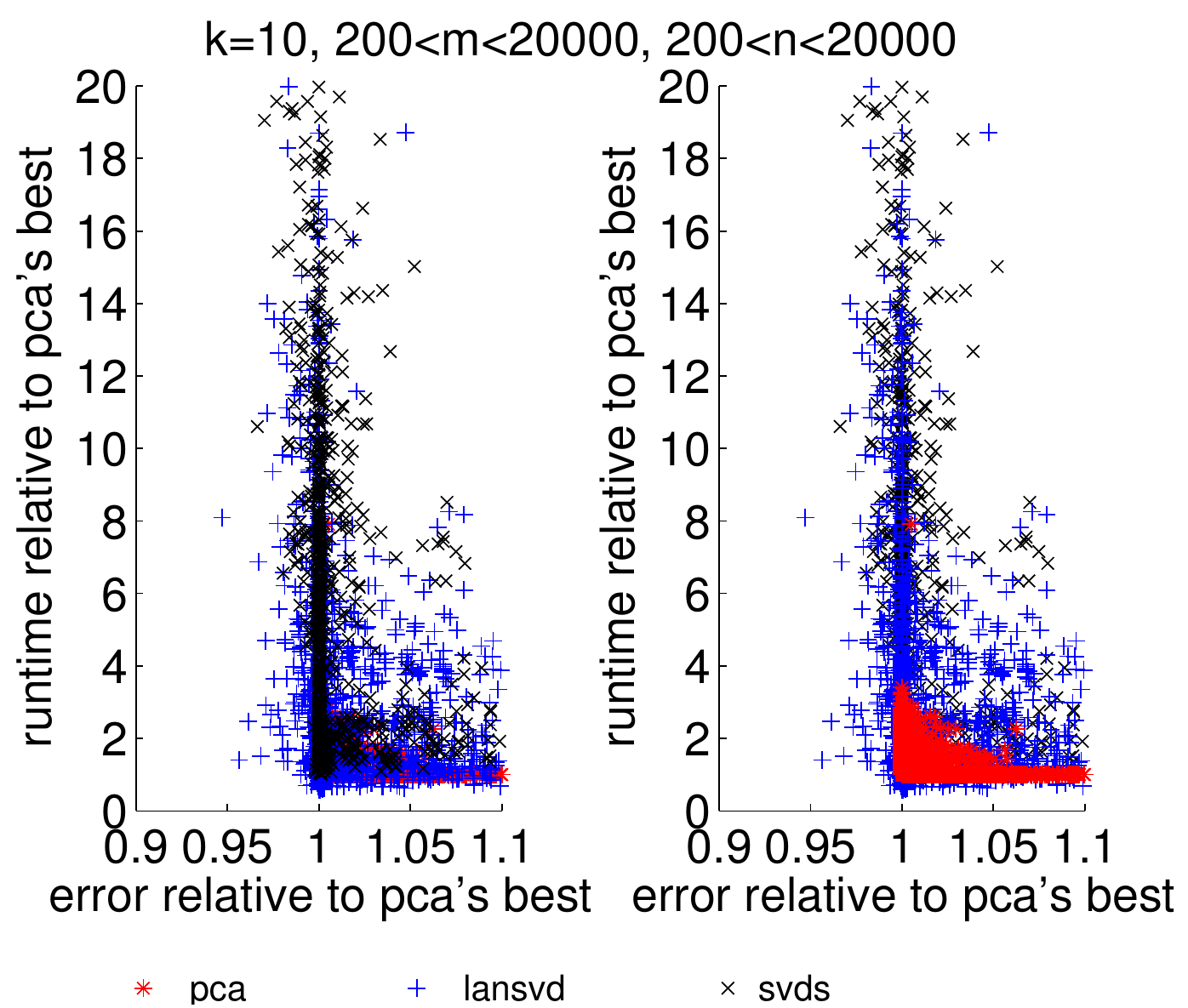}}
\parbox{.49\textwidth}{\includegraphics[width=.48\textwidth]{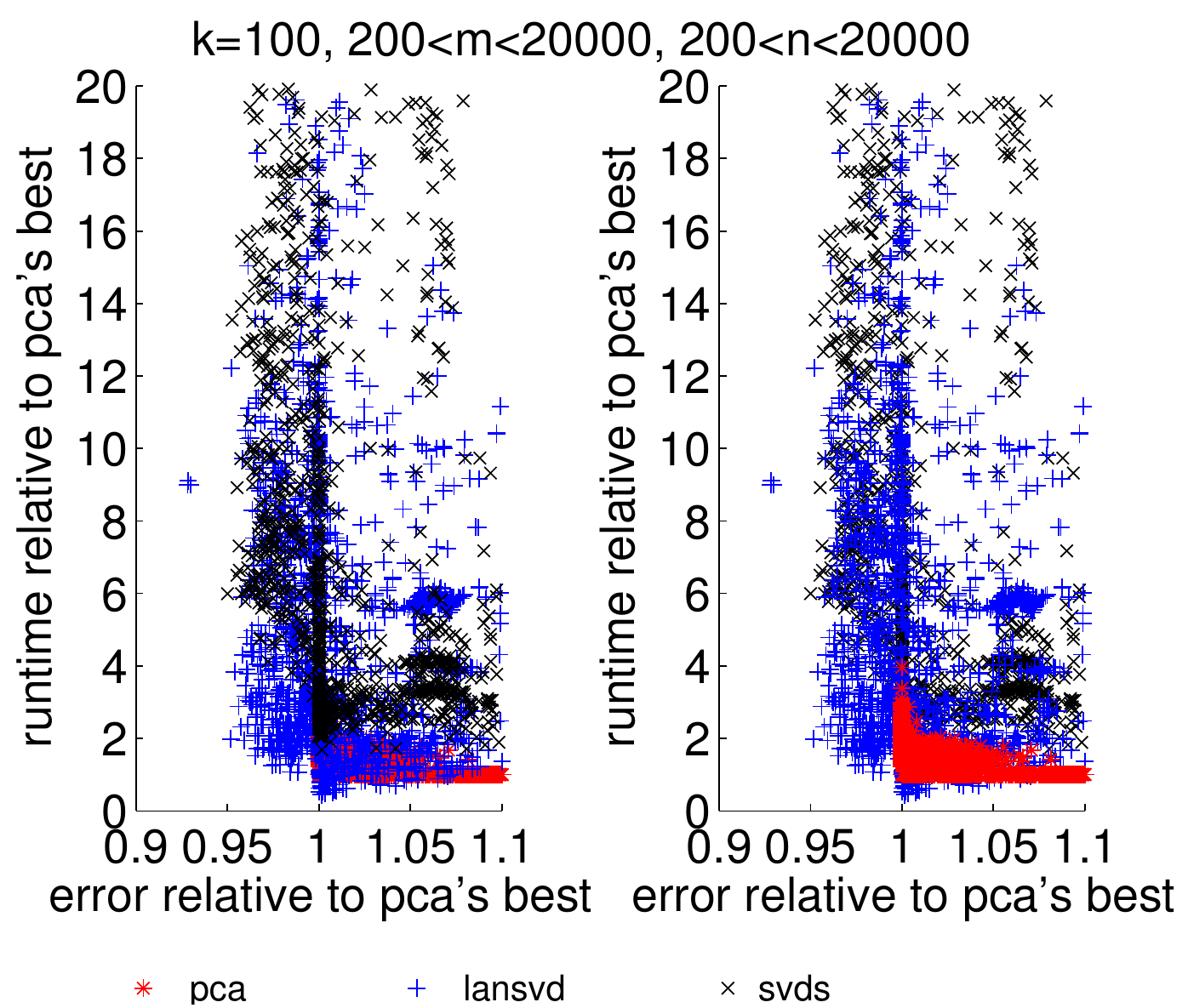}}
\\\ \\
\caption{svds vs.\ lansvd vs.\ pca for small sparse matrices}
\label{svdsfig}
\end{figure}

For each of the two parameter settings displayed in Figure~\ref{svdsfig}
(namely, $k=10$ and $k=100$), we plot the spectral-norm errors and runtimes
for pca (our code), lansvd (PROPACK of~\citet{larsen}), and MATLAB's built-in
svds (ARPACK of~\citet{arpack}). For pca, we vary the parameter specifying the
number of iterations, its $= 2$, 5, 8; we leave the oversampling parameter $l$
that specifies the number of random vectors whose entries are i.i.d.\ at the
default, $l = k+2$. For lansvd, we vary the tolerance for convergence as
tol $= 10^{-5}$, 1, $10^3$. For svds, we vary the tolerance as for lansvd,
but with $10^{-6}$ in place of $10^{-5}$ (svds requires a tighter tolerance
than lansvd to attain the best accuracy). The red asterisks correspond to pca,
the blue ``plus signs'' correspond to lansvd, and the black ``times signs''
correspond to svds. Please note that lansvd may return
entirely erroneous results without warning, as indicated
in Section~\ref{proprob}. The plotted accuracy is the spectral norm
of the difference between $A$ and the computed rank-$k$ approximation;
pca's error was always at most twice the best for any of the three algorithms
for any setting of parameters.
Each plot in Figure~\ref{svdsfig} appears twice,
once with svds on top of lansvd on top of pca,
and once with pca on top of lansvd on top of svds.
In Figure~\ref{svdsfig}, pca generally exhibits higher performance than svds.

The computations used MATLAB 8.3.0.532 (R2014a) on a four-processor machine,
with each processor being an Intel Xeon E5-2680 v2 containing 10 cores,
where each core operated at 2.8\,GHz, with 25.6\,MB of L2 cache.
We did not consider the MATLAB Statistics Toolbox's own ``pca,'' ``princomp,''
and ``pcacov,'' as these compute all singular values and singular vectors
(not only those relevant for low-rank approximation), just like MATLAB's
built-in svd (in fact, these other functions call svd).

\section{Conclusion}
\label{conclusion}

On strictly serial processors with no complicated caching (such as the
processors of many decades ago), the most careful implementations of Lanczos
iterations by~\citet{larsen} and others could likely attain performance nearing
the randomized methods', unlike competing techniques such as the power method
or the closely related nonlinear iterative partial least squares (NIPALS)
of~\citet{wold}. The randomized methods can attain much higher performance on
parallel and distributed processors, and generally are easier to use ---
setting their parameters properly is trivial (defaults are fine), in marked
contrast to the wide variance in performance of the classical schemes with
respect to inputs and parameter settings. Furthermore, despite decades of
research on Lanczos methods, the theory for the randomized algorithms is more
complete and provides strong guarantees of excellent accuracy, whether or not
there exist any gaps between the singular values. With regard to principal
component analysis for low-rank approximation, Lanczos iterations are like
complicated, inherently serial heuristics for trying to emulate the reliable
and more easily parallelized randomized methods. The randomized algorithms
probably should be the methods of choice for computing the low-rank
approximations in principal component analysis, when implemented and validated
with consideration for the developments
in Sections~\ref{nystrom}--\ref{proprob} above.

\section*{SOFTWARE}

Our MATLAB implementation is available at http://tygert.com/software.html

\bibliography{bench}
\bibliographystyle{ACM-Reference-Format-Journals}

\end{document}